\documentclass[twocolumn]{aastex631}
\NewPageAfterKeywords
\usepackage{comment}
\usepackage{xstring}
\providecommand{\sci}[1]{\protect\ensuremath{\times 10^{\StrSubstitute[0]{#1}{e}{}}}}
\usepackage[flushmargin]{footmisc}

\newcommand{\pokemonnum}{1125}
\newcommand{\primaryfifteen}{455}
\newcommand{\singlefifteen}{348}
\newcommand{\multiplefifteen}{107}
\newcommand{\doublefifteen}{85}
\newcommand{\triplefifteen}{21}
\newcommand{\quadruplefifteen}{0}
\newcommand{\quintuplefifteen}{1}
\newcommand{\multiplicityrate}{$23.5\pm2.0\%$} 
\newcommand{\companionrate}{$28.8\pm2.1\%$} 
\newcommand{\noplanetpeak}{5.57}
\newcommand{\planetpeak}{198}
\newcommand{\toipeak}{735}

\received{September 22, 2023}
\revised{January 6, 2024}
\accepted{January 24, 2024}

\submitjournal{AJ}

\graphicspath{{./}{figures/}}

\begin{document}

\title{The POKEMON Speckle Survey of Nearby M Dwarfs. III. The Stellar Multiplicity Rate of M Dwarfs within 15 pc}

\correspondingauthor{Catherine A. Clark}
\email{catherine.a.clark@jpl.nasa.gov}

\author[0000-0002-2361-5812]{Catherine A. Clark}
\affil{Jet Propulsion Laboratory, California Institute of Technology, Pasadena, CA 91109 USA}
\affil{NASA Exoplanet Science Institute, IPAC, California Institute of Technology, Pasadena, CA 91125 USA}

\author[0000-0002-8552-158X]{Gerard T. van Belle}
\affil{Lowell Observatory, 1400 West Mars Hill Road, Flagstaff, AZ 86001, USA}

\author[0000-0003-2159-1463]{Elliott P. Horch}
\affil{Southern Connecticut State University, 501 Crescent Street, New Haven, CT 06515, USA}

\author[0000-0002-5741-3047]{David R. Ciardi}
\affil{NASA Exoplanet Science Institute, IPAC, California Institute of Technology, Pasadena, CA 91125 USA}

\author[0000-0002-5823-4630]{Kaspar von Braun}
\affil{Lowell Observatory, 1400 West Mars Hill Road, Flagstaff, AZ 86001, USA}

\author[0000-0001-5306-6220]{Brian A. Skiff}
\affil{Lowell Observatory, 1400 West Mars Hill Road, Flagstaff, AZ 86001, USA}

\author[0000-0001-6031-9513]{Jennifer G.\ Winters}
\affil{Bridgewater State University, 131 Summer St., Bridgewater, MA 02324, USA}
\affil{Center for Astrophysics $\vert$ Harvard \& Smithsonian, 60 Garden Street, Cambridge, MA 02138, USA}

\author[0000-0003-2527-1598]{Michael B. Lund}
\affil{NASA Exoplanet Science Institute, IPAC, California Institute of Technology, Pasadena, CA 91125 USA}

\author[0000-0002-0885-7215]{Mark E. Everett}
\affiliation{NSF’s National Optical-Infrared Astronomy Research Laboratory, 950 N. Cherry Ave., Tucson, AZ 85719, USA}

\author[0000-0003-4236-6927]{Zachary D. Hartman}
\affil{Gemini Observatory/NSF's NOIRLab, 670 A'ohoku Place, Hilo, HI 96720 USA}

\author[0000-0003-4450-0368]{Joe Llama}
\affil{Lowell Observatory, 1400 West Mars Hill Road, Flagstaff, AZ 86001, USA}



\begin{abstract}

M dwarfs are ubiquitous in our galaxy, and the rate at which they host stellar companions, and the properties of these companions, provide a window into the formation and evolution of the star(s), and of any planets that they may host. The Pervasive Overview of `Kompanions' of Every M dwarf in Our Neighborhood (POKEMON) speckle survey of nearby M dwarfs is volume-limited from M0V through M9V out to 15 pc, with additional targets at larger distances. In total, \pokemonnum{} stars were observed, and \primaryfifteen{} of these are within the volume-limited, 15-pc sample of M-dwarf primaries. When we combine the speckle observations with known companions from the literature, we find that the stellar multiplicity rate of M dwarfs within 15 pc is \multiplicityrate{}, and that the companion rate is \companionrate{}. We also find that the projected separation distribution for multiples that are known to host planets peaks at \planetpeak{} au, while the distribution for multiples that are not yet known to host planets peaks at \noplanetpeak{} au. This result suggests that the presence of close-in stellar companions inhibits the formation of M-dwarf planetary systems, similar to what has been found for FGK stars.

\end{abstract}



\section{Introduction} \label{sec:introduction}

Due to their small sizes and low temperatures and luminosities, the M dwarfs currently have the most favorable observing characteristics for finding Earth-sized planets in the habitable zone via the radial velocity and transit methods \citep[e.g.,][]{Shields2016PhR...663....1S}. The M dwarfs also comprise over 70\% of the stars in the Milky Way \citep{Henry2006AJ....132.2360H, Winters2015AJ....149....5W}, and have lifetimes on the order of the age of the Universe \citep{Laughlin1997ApJ...482..420L}, giving planets a place and the time to form and evolve.


However, it has been shown that the presence of a stellar companion can impact or inhibit the formation and evolution of planetary systems. In particular, close-in stellar companions can dynamically impact the evolution of protoplanetary disks and planetesimals during the early stages of planet formation \citep{DesideraBarbieri2007A&A...462..345D, HaghighipourRaymond2007ApJ...666..436H, Jang-Condell2015ApJ...799..147J, RafikovSilsbee2015ApJ...798...69R, RafikovSilsbee2015ApJ...798...70R}. However, recent studies have suggested that even wide stellar companions might affect the formation or orbital properties of giant planets \citep{Fontanive2019MNRAS.485.4967F, FontaniveGagliuffi2021FrASS...8...16F, Hirsch2021AJ....161..134H, Su2021AJ....162..272S, Mustill2022A&A...658A.199M}.

Additionally, stellar companions inhibit our ability to detect and characterize planets once formed. Stellar companions can induce ``third light'' contamination into light curves from missions such as Kepler \citep{Borucki2011ApJ...728..117B}, K2 \citep{Howell2014PASP..126..398H}, and the Transiting Exoplanet Survey Satellite \citep[TESS;][]{Ricker2015JATIS...1a4003R}, which can inhibit the detection of Earth-sized planets via the transit method \citep{Lester2021AJ....162...75L}, as well as the characterization of planets in general \citep{Hirsch2017AJ....153..117H, Bouma2018AJ....155..244B, Teske2018AJ....156..292T, FurlanHowell2020ApJ...898...47F, Howell2020FrASS...7...10H}. Stellar companions can also inhibit our ability to detect planets via the radial velocity method if the signal from the companion is not taken into account \citep[e.g.,][]{Ortiz2016A&A...595A..55O}. Furthermore, any statistical value for $\eta_{\oplus}$ will be affected by unseen companions, as actual planet radii are larger than observed radii in multi-star systems \citep{Ciardi2015ApJ...805...16C}.

In addition to impacting observed planetary properties, stellar companions may also affect the physical properties of any planets in the multi-star system, such as their orbits and masses \citep{Eggenberger2004A&A...417..353E}.

As the M dwarfs offer such a significant opportunity for finding and characterizing Earth-sized planets in the habitable zone \citep[e.g.,][]{Lopez-Morales2019AJ....158...24L}, understanding the rate at which the M dwarfs host stellar companions -- and the properties of these companions -- is crucial to the field of exoplanet studies.

For this reason, a number of studies have been carried out to measure the multiplicity rate of exoplanet host stars \citep[e.g.,][and subsequent papers in this series]{MugrauerMichel2020AN....341..996M}, as well as the multiplicity rate of field M dwarfs (Table \ref{table:previous_studies}). However, many of these studies were limited by the size of their sample, due to the fact that the M dwarfs are quite faint in the optical, and that a reasonably-complete inventory of later M dwarfs did not even exist until recently \citep{Kirkpatrick2014ApJ...783..122K, LuhmanSheppard2014ApJ...787..126L, Winters2021AJ....161...63W}. Additionally, many of these studies were not sensitive to close-in stellar companions due to the resolution limits of the surveys.

\startlongtable
\begin{deluxetable}{cc}
\tablecaption{Previous studies of M-dwarf multiplicity
\label{table:previous_studies}}
\tablehead{\colhead{Reference} & \colhead{Multiplicity Rate} \\
\colhead{} & \colhead{(\%)}}
\startdata
\citet{Henry1991PhDT........11H} & 30-40 \\
\citet{FischerMarcy1992ApJ...396..178F} & $42\pm9$ \\
\citet{Simons1996AJ....112.2238S} & 40 \\
\citet{Leinert1997AA...325..159L} & $26\pm9$ \\
\citet{ReidGizis1997AJ....113.2246R} & 35 \\
\citet{Law2008MNRAS.384..150L} & $13.6^{+6.5}_{-4.0}$ \\
\citet{Bergfors2010AA...520A..54B} & $32\pm6$ \\
\citet{Janson2012ApJ...754...44J} & $27\pm3$ \\
\citet{Jodar2013MNRAS.429..859J} & $20.3^{+6.9}_{-5.2}$ \\
\citet{Janson2014ApJ...789..102J} & 21-27 \\
\citet{Ward-Duong2015MNRAS.449.2618W} & $23.5\pm3.2$ \\
\citet{Cortes-Contreras2017AA...597A..47C} & $19.5\pm2.3$ \\
\citet{Winters2019AJ....157..216W} & $27.5\pm1.4$ \\
\hline
This Work & \multiplicityrate{} \\
\enddata
\end{deluxetable}

We have therefore carried out the ``Pervasive Overview of Kompanions of Every M dwarf in Our Neighborhood'' (POKEMON) speckle survey of nearby M dwarfs. The POKEMON survey was designed to be volume-limited from M0V through M9V and out to 15 pc, with additional brighter targets at larger distances. This volume-limited nature of the sample allowed for uniform sampling of close-in stellar companions. Additionally, since the targets are nearby, the spatial resolution into the systems is improved. In total, \pokemonnum{} targets were observed at diffraction-limited resolution. The first paper in the series presented the new discoveries \citep{Clark2022AJ....164...33C}, and the second paper in the series presented all companions detected within our speckle images \citep{Clark2024AJ....167...56C}. In this third paper in the series, we refine the assessment of the stellar multiplicity rate of M dwarfs within 15 pc.

In Section \ref{sec:sample}, we define the 15-pc POKEMON sample of M-dwarf primaries. In Section \ref{sec:multiplicity_rate}, we present the stellar companions to the targets in our 15-pc sample that are known to the literature. In Section \ref{sec:stellar_multiple_gap}, we describe the observed gap between the projected separation distributions for planet-hosting and non-planet-hosting multiples. Finally, in Section \ref{sec:conclusions}, we summarize our conclusions and detail the additional work on the POKEMON sample that will be carried out in the future.

\section{Definition of the Sample} \label{sec:sample}

The creation of the POKEMON sample as a whole has been described extensively in the previous two papers in the series \citep{Clark2022AJ....164...33C, Clark2024AJ....167...56C}. Briefly, the POKEMON sample was designed to be volume-limited through M9V out to 15 pc, though it also includes additional brighter targets at larger distances for a total sample size of \pokemonnum{} targets. At the time the POKEMON catalog was constructed (circa 2015), the generally accepted cutoff for the stellar fusion-burning limit was M9V; however, \citet{Dieterich2014AJ....147...94D} defined the barrier between the lowest-mass stars and the brown dwarfs at a spectral type of L2V. We continue to use the former definition in order to establish the stellar multiplicity of the M-dwarf sequence in particular.

In order to determine the stellar multiplicity and companion rates of the M dwarfs from a volume-limited sample, we needed to constrain the POKEMON targets to only include only those within 15 pc. The volume-limited nature of the sample allows for uniform sampling of close-in stellar companions amongst our low-mass neighbors.

We used Gaia Data Release 3 \citep[DR3;][]{Gaia2023AA...674A...1G} parallaxes, as well as parallaxes from other literature sources when Gaia DR3 data were not available, to define the the 15-pc POKEMON sample. 70 of the \pokemonnum{} objects in the original POKEMON sample do not have parallaxes in Gaia DR3; we obtained parallaxes from 18 additional sources to determine whether or not they lie within 15 pc. We were unable to obtain distance measurements for two of these objects.

We note that the initial POKEMON catalog was developed before data from Gaia were available. When the parallax sources for the POKEMON targets were updated using the Gaia astrometric data, some sources left the 15-pc sample, and we did not add additional sources into the sample. This represents a source of incompleteness, but there was no bias in how the sources were removed from the volume. The sample is therefore smaller – meaning the error bars on our multiplicity and companion rates are larger – but it remains unbiased.

In any case, we find that the parallax sources used to create the POKEMON catalog are mostly consistent with Gaia; a cursory examination of Gaia DR3 indicates 606 northern hemisphere ($\delta > -30^o$) late-type dwarfs \citep[based on a  $B_p-R_p > 1.80$ cut to select M-dwarfs as suggested by][]{PecautMamajek2013ApJS..208....9P} within 15 pc. Although this value is larger than the number of targets in the 15-pc POKEMON sample described in this paper, we note that some of the targets in the Gaia sample are secondaries and not primaries (82), brown dwarfs (46), white dwarfs (3), or RR Lyrae variables (1). This means that 474 are M-dwarf primaries. Of these, 403 are within the 15-pc POKEMON sample. As noted in \citet{Clark2024AJ....167...56C}, Gaia is frequently unable to provide an astrometric solution (including a parallax) for close binaries. As such, the 15-pc POKEMON sample includes 52 targets that were not identified by Gaia as late-type dwarfs within 15 pc. Nonetheless, in the future we plan to present an expanded version of the POKEMON catalog that includes the Gaia late-type dwarfs within 15 pc that we have not yet observed as a part of the POKEMON survey\footnote{The POKEMON Distance-Limited Catalog (POKEMON-DLC)}.

\begin{figure*}
    \centering
    \includegraphics[width=\textwidth]{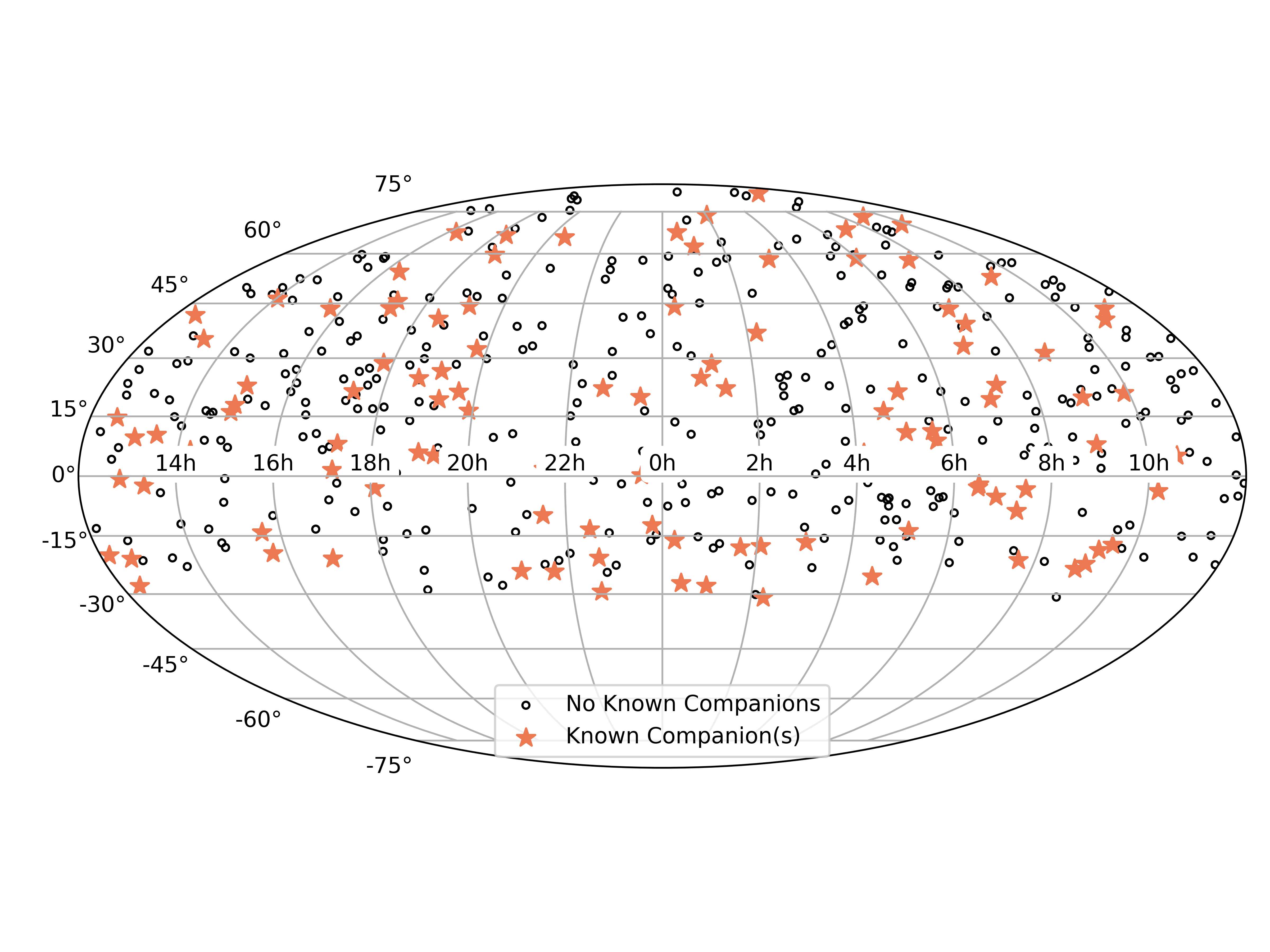}
    \caption{The sky locations of the \primaryfifteen{} targets in the volume-limited, 15-pc POKEMON sample. Targets with no known companions are marked with open black circles, while targets with known companions are marked with larger, orange stars.}
    \label{fig:aitoff}
\end{figure*}

\begin{figure*}
    \centering
    \includegraphics[width=0.49\textwidth]{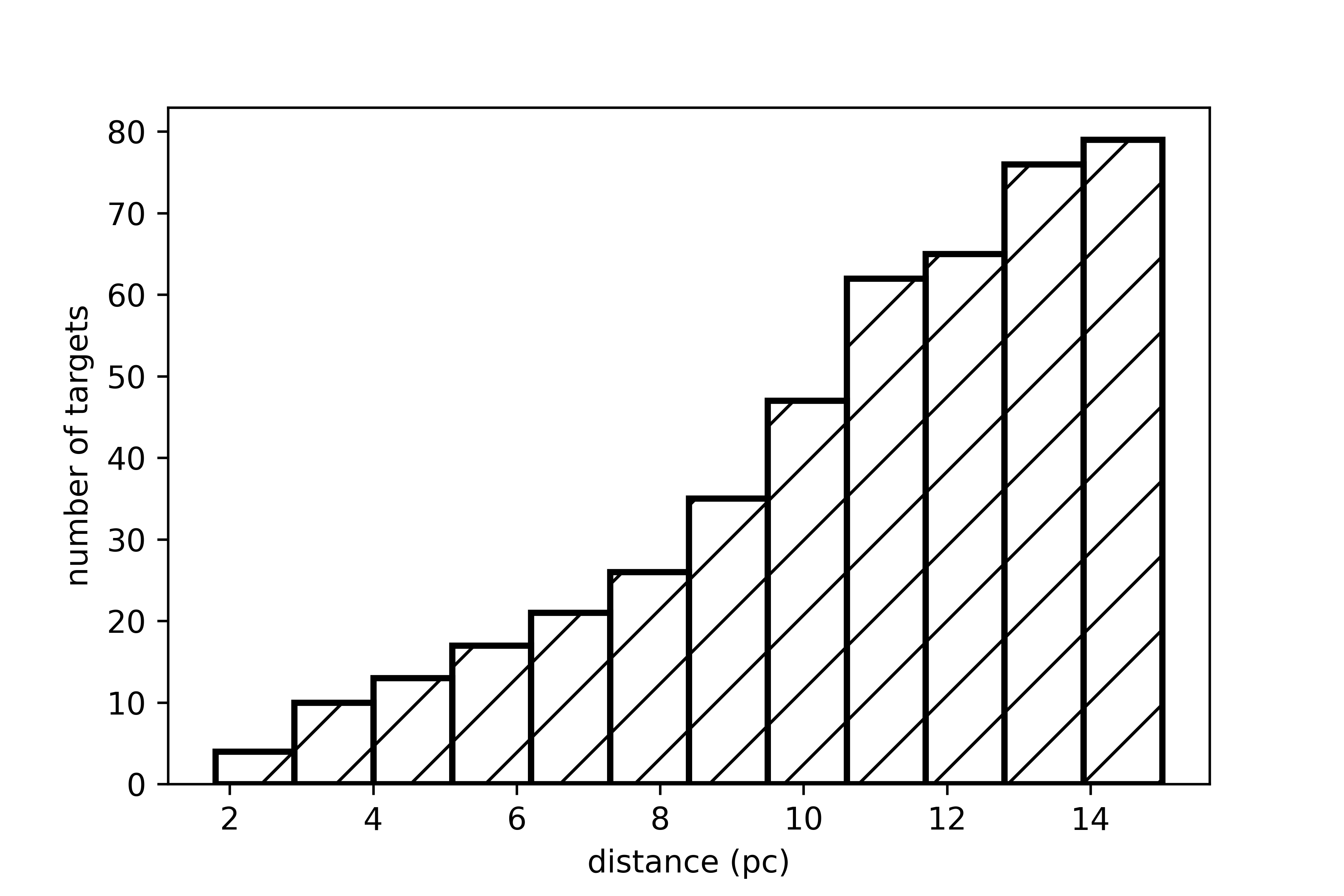}
     \includegraphics[width=0.49\textwidth]{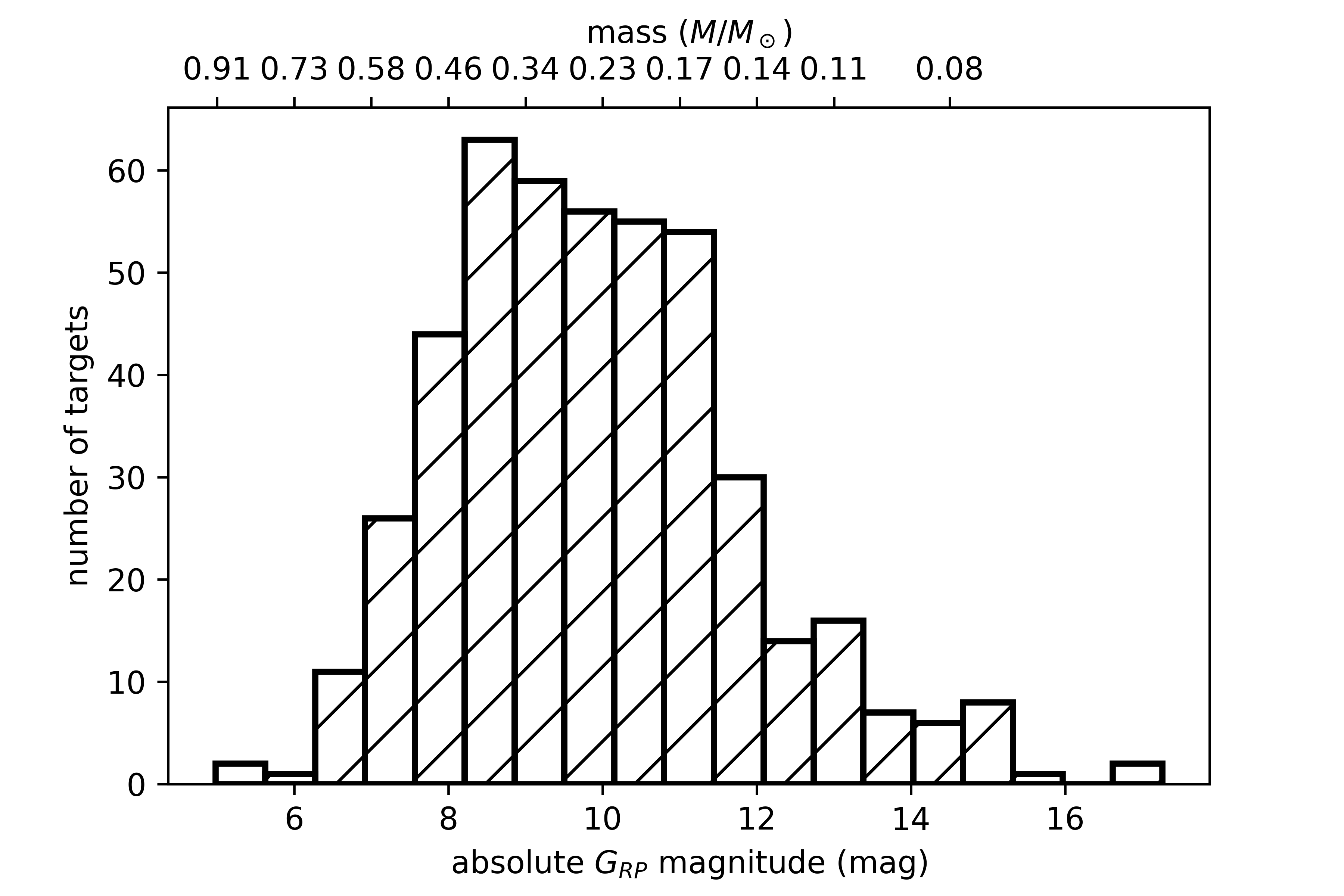}
    \caption{Distance (left) and absolute $G_{RP}$ magnitude (right) distributions for the \primaryfifteen{} targets in the 15-pc POKEMON sample. We use the mass-magnitude relation from \citet{GiovinazziBlake2022AJ....164..164G} to estimate the masses of these targets.}
    \label{fig:histograms}
\end{figure*}

In total, the volume-limited, 15-pc POKEMON sample of M-dwarf primaries consists of \primaryfifteen{} targets (Figure \ref{fig:aitoff}). This sample is characterized in Table \ref{table:targets}, where we include the Two Micron All Sky Survey \citep[2MASS;][]{Skrutskie2006AJ....131.1163S} ID or name, Gaia DR3 ID, Gaia $G$ magnitude, Gaia $G_{RP}$ magnitude, parallax-derived distance, and reference for the distance. We also note whether the target has any known companions. In Figure \ref{fig:histograms} we show histograms of the distances and absolute $G_{RP}$ magnitudes for the targets in the POKEMON sample. We use the absolute $G_{RP}$ magnitudes and the mass-magnitude relation from \citet{GiovinazziBlake2022AJ....164..164G} to estimate masses for the POKEMON targets.

\begin{deluxetable*}{ccccccc}
\tablecaption{Targets in the volume-limited, 15-pc POKEMON sample of M-dwarf primaries
\label{table:targets}}
\tablehead{\colhead{2MASS ID or Name} & \colhead{Gaia DR3 ID} & \colhead{$G$ Magnitude} & \colhead{$G_{RP}$ Magnitude} & \colhead{Distance} & \colhead{Reference} & \colhead{Known companion(s)?} \\ 
\colhead{} & \colhead{} & \colhead{(mag)} & \colhead{(mag)} & \colhead{(pc)} & \colhead{} & \colhead{}}
\rotate
\startdata
00064325-0732147 & 2441630500517079808 & 11.8 & 10.4 & 4.8 & 6 & N \\
00085512+4918561 & 393621524910343296 & 14.4 & 13 & 14.8 & 6 & N \\
00113182+5908400 & 423027104407576576 & 13.5 & 12.1 & 9.3 & 6 & N \\
00152799-1608008 & 2368293487261055488 & 10.3 & 8.9 & 5.0 & 10 & Y \\
00153905+4735220 & 392350008432819328 & 10 & 8.9 & 11.5 & 6 & N \\
00154919+1333218 & 2768048564768256512 & 11.4 & 10.2 & 12.2 & 6 & N \\
00182256+4401222 & 385334230892516480 & 7.2 & 6.2 & 3.6 & 6 & Y \\
00202922+3305081 & 2863419584886542080 & 13.9 & 12.5 & 12.3 & 6 & N \\
00242463-0158201 & 2541756977144595712 & 16.6 & 15 & 12.4 & 6 & N \\
00244419-2708242 & 2322561156529549440 & 13.2 & 11.6 & 7.7 & 6 & Y \\
\enddata
\tablecomments{Table \ref{table:targets} is published in its entirety in the machine-readable format. A portion is shown here for guidance regarding its form and content.}
\tablerefs{(1) \citet{Dittmann2014ApJ...784..156D}; (2) \citet{Dupuy2019AJ....158..174D}; (3) \citet{DupuyLiu2017ApJS..231...15D}; (4) \citet{Finch2018AJ....155..176F}; (5) \citet{Gaia2018AA...616A...1G}; (6) \citet{Gaia2023AA...674A...1G}; (7) \citet{Henry2006AJ....132.2360H}; (8)  \citet{Riedel2010AJ....140..897R}; (9) \citet{Torres2010AARv..18...67T}; (10) \citet{vanLeeuwen2007AA...474..653V}.}
\end{deluxetable*}

We note that most speckle cameras have a faint limit at $I=15.5$ due to the short exposure times, although by design this magnitude cutoff was expected to be sufficiently faint to include M9V stars out to 15 pc. \citet{Kirkpatrick1991ApJS...77..417K} lists LHS 2924 as the primary M9V spectral standard, and LHS 2065 as the secondary M9V spectral standard. LHS 2924 \citep[$I_{c}$ = 15.2; $\varpi = 91.2\pm0.1$ mas;][]{Dahn2017AJ....154..147D, Gaia2023AA...674A...1G} and LHS 2065 \citep[$I_{c}$ = 14.4; $\varpi = 115.49\pm0.07$ mas;][]{Leggett2001ApJ...548..908L, Gaia2023AA...674A...1G} have absolute magnitudes of $M_{Ic}$ = 15.0 and 14.8, respectively. Two other well-characterized, nearby ($d < 11$ pc) M9V stars -- LP 944-20 \citep[$I_{c}$ = 14.0; $\varpi = 155.60\pm0.05$ mas; $M_{Ic}$ = 15.0;][]{Dieterich2014AJ....147...94D, Gaia2023AA...674A...1G, Kirkpatrick1999ApJ...519..802K, Henry2004AJ....128.2460H} and LP 647-13 \citep[$I_{c}$ = 14.9; $\varpi = 94.6\pm0.2$ mas; $M_{Ic}$ = 14.7;][]{Gaia2023AA...674A...1G, Cruz2003AJ....126.2421C} -- have similar absolute magnitudes. We therefore adopt the average absolute magnitude of the two standards LHS 2924 and LHS 2065 ($M_{Ic}$ $\simeq$ 14.9) as typical for field M9V stars (E. Mamajek, private communication). Setting $M_I=14.9$ and $d=15$, and solving

\begin{equation}
    m_I - M_I = -5 + 5\log_{10}(d)
\end{equation}

\noindent for $m_I$, we find that we would need to observe targets as faint as $m_I=15.8$ in order to observe M9V stars out to 15 pc. The Gaia $G_{RP}$ filter has a central wavelength at 797 nm, which is comparable to the $I$-band central wavelength at 806 nm. The faintest target in the 15-pc POKEMON sample has an apparent $G_{RP}$ magnitude of 17.1, which is sufficiently faint to detect an M9V star at 15 pc.

Using the apparent $G$ magnitudes from Gaia, the stellar distances, and the reference stellar properties from \citet{PecautMamajek2013ApJS..208....9P}, we estimate spectral types for the POKEMON targets (Table \ref{table:spectral_types}). It is important to note that the original POKEMON sample was defined based on spectral types derived from colors; therefore, some of the spectral types estimated using the apparent $G$ magnitudes are either K or L dwarfs. The fourth paper in the series will present a homogenous set of photometrically-derived spectral types for the POKEMON targets, allowing for more accurate stellar characterization, and establishing the M-dwarf multiplicity rate by spectral subtype through M9V for the first time.

\startlongtable
\begin{deluxetable}{cc}
\tablecaption{Spectral types of the targets in the 15-pc POKEMON sample
\label{table:spectral_types}}
\tablehead{\colhead{Spectral Type} & \colhead{Number of Targets}}
\startdata
K & 26 \\
M0 & 33 \\
M1 & 35 \\
M2 & 57 \\
M3 & 115 \\
M4 & 92 \\
M5 & 69 \\
M6 & 5 \\
M7 & 8 \\
M8 & 6 \\
M9 & 1 \\
L & 8 \\
\enddata
\end{deluxetable}

\section{The Stellar Multiplicity Rate of M Dwarfs Within 15 pc} \label{sec:multiplicity_rate}

\begin{figure*}
    \centering
    \includegraphics[width=\textwidth]{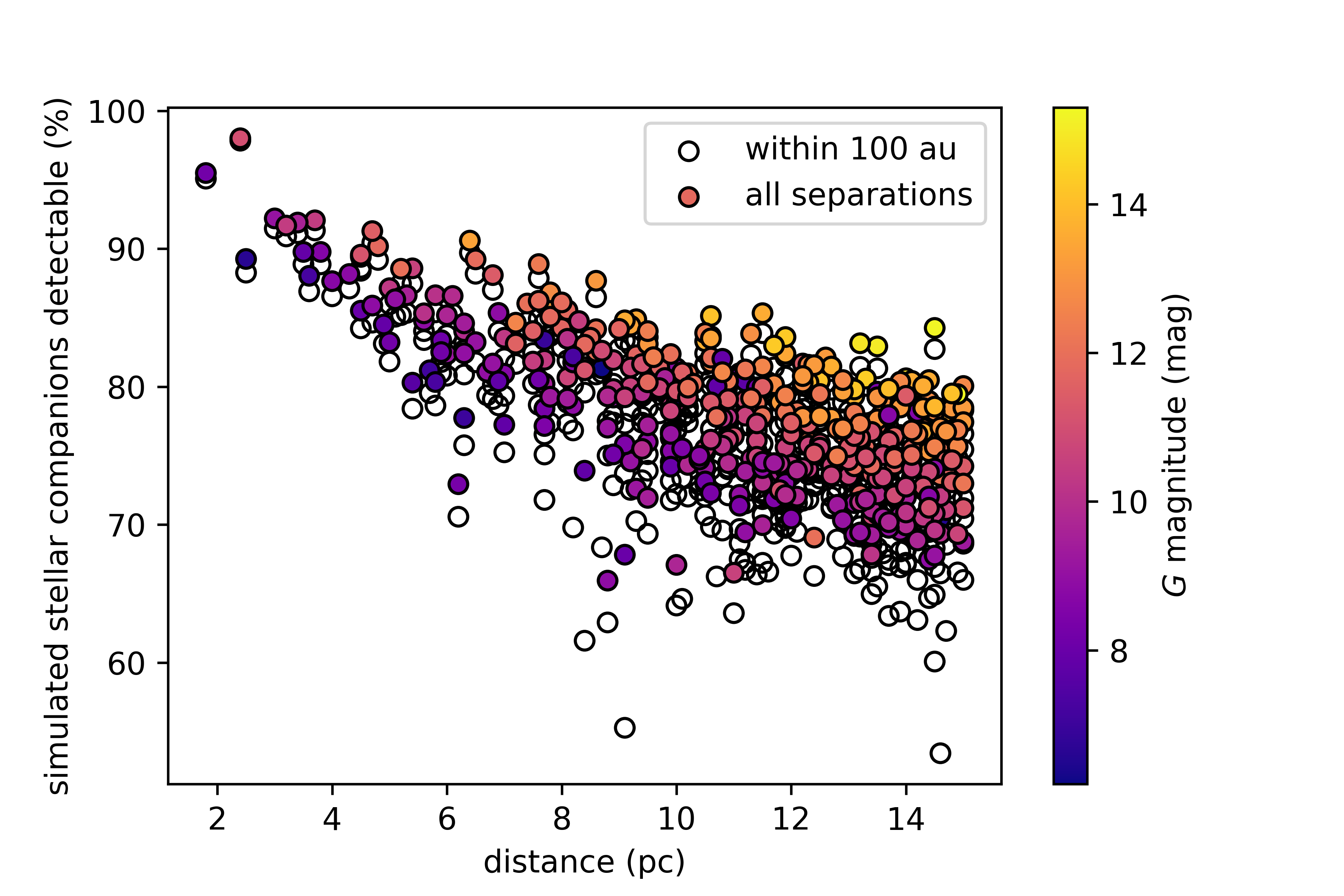}
    \caption{Percentage of stellar companions detectable within our speckle images versus distance. Circles color-coded by $G$ magnitude represent the detectability of simulated companions at all separations. Open black circles represent the detectability of simulated companions specifically within 100 au. We find that the majority of potential stellar companions would be detectable, particularly for the nearby targets, with a median value of 78.1\%. Specifically within 100 au, we find that 75.5\% of simulated companions are recovered.}
    \label{fig:percent_detectable_vs_distance}    
\end{figure*}

As noted in \citet{Winters2019AJ....157..216W}, most stellar companions to nearby M dwarfs are close-in ($<50$ au), and speckle interferometry -- the technique used to image the targets in the POKEMON sample -- is uniquely suited to detecting close-in stellar companions. In Figure \ref{fig:percent_detectable_vs_distance}, we plot the percent of simulated stellar companions detectable within our speckle images as a function of distance. This analysis was carried out following the procedure outlined in Section 4.2 of the previous paper in the series \citep{Clark2024AJ....167...56C}. Briefly, the code works by first identifying the population of stellar companions that could orbit each star using the Dartmouth isochrones \citep{Dotter2008ApJS..178...89D} and the \citet{DucheneKraus2013ARA&A..51..269D} mass ratio and orbital period distributions for M dwarfs, and then uses the derived contrast curves to evaluate the sensitivity of each observation to these simulated companions. This plot shows that speckle imaging would allow us to recover the majority (78.1\%) of simulated stellar companions to the targets in the 15-pc POKEMON sample. For simulated companions specifically within 100 au, we recover 75.5\%.

Nonetheless, every detection technique only fills one particular gap in parameter space. Since the POKEMON survey used only one technique -- speckle interferometry -- to detect stellar companions, we carried out a literature search to ensure the inclusion of all known stellar companions to the M-dwarf primaries in our 15-pc sample. Our accompanying literature search ensured that wider $(\gtrapprox2\arcsec)$ companions were not overlooked, and ensured the accuracy of our multiplicity and companion rate calculations.

For the literature search, we mainly used data obtained via the Washington Double Star catalog \citep[WDS;][]{Mason2001AJ....122.3466M}, and the Simbad bibliography tool \citep{Wenger2000A&AS..143....9W}. However, Simbad and the WDS do not provide a complete census of known companions, so we also carried out our own literature search to search for additional companions, and to ensure that the WDS companions are bound. We also used the SUPERWIDE catalog \citep{HartmanLepine2020ApJS..247...66H} to identify any wide, common-proper-motion pairs.

In Table \ref{table:known_companions}, we report the properties of the known stellar companions to the POKEMON primaries within 15 pc that were identified either from our speckle survey or the literature search. These are M-dwarf companions to the M-dwarf primaries in the sample; as we are calculating the stellar multiplicity rate for M dwarfs, we do not consider any substellar companions such as brown dwarfs or planets. We also do not consider M dwarfs that have more massive companions, such as higher-mass Main Sequence stars, evolved stars, or white dwarfs.

Table \ref{table:known_companions} includes the 2MASS ID or name of the primary, component configuration for each row, year of the measurement, position angle ($\theta$), angular separation ($\rho$), projected separation ($a$), detection technique, reference for the detection technique, magnitude difference ($\Delta m$), filter of the magnitude difference, and reference for the magnitude difference. We also list uncertainties on position angle, angular separation, and magnitude difference when available.

There are $n-1$ lines for each system, where $n$ is the number of components in the system. For instance, the quintuple system 2MASS J16552880-0820103 has four lines in Table \ref{table:known_companions}. For the higher-order multiples, the component configuration indicates the components used to measure the listed magnitude difference. For instance, in the case of the 2MASS J00244419-2708242 triple system, the first line lists the magnitude difference between the primary and the photocenter of the companions (A-BaBb), while the second line lists the magnitude difference between the two stars that make up the B component (Ba-Bb).

For the astrometric and spectroscopic binaries, the projected separations listed in Table \ref{table:known_companions} were either taken from the reference given, or were calculated from the orbital period and the estimated masses of the components via Kepler's third law. We were able to calculate a projected separation for all multiples other than 2MASS J19352922+3746082 -- which was first identified as a radial velocity variable by \citet{Shkolnik2012ApJ...758...56S} and confirmed as an SB1 by \citet{Jeffers2018AA...614A..76J} -- as no orbital elements or stellar properties have been published in the literature.

\begin{deluxetable*}{cccccccccccccc}
\tablecaption{Summary of known stellar companions to POKEMON primaries within 15 pc}
\label{table:known_companions}
\tablehead{\colhead{ID} & \colhead{Component} & \colhead{Epoch} & \colhead{$\theta$} & \colhead{Error} & \colhead{$\rho$} & \colhead{Error} & \colhead{$a$} & \colhead{Technique} & \colhead{Reference} & \colhead{$\Delta m$} & \colhead{Error} & \colhead{Filter} & \colhead{Reference} \\ 
\colhead{} & \colhead{} & \colhead{(yr)} & \colhead{($^{\circ}$)} & \colhead{($^{\circ}$)} & \colhead{($\arcsec$)} & \colhead{($\arcsec$)} & \colhead{(au)} & \colhead{} & \colhead{} & \colhead{(mag)} & \colhead{(mag)} & \colhead{} & \colhead{}}
\rotate
\startdata
00152799-1608008 & A-B & 1985 & 130 &  & 0.22 &  & 1.0945 & micdet & 30 & 0.4 &  & V & 30 \\
00182256+4401222 & A-B & 1907 & 57 &  & 38.83 &  & 138.33 & micdet & 11 & 3.02 &  & V & 88 \\
00244419-2708242 & A-BaBb & 1993 & 353 & 0.4 & 1.07 & 0.4 & 8.2743 & spkdet & 55 & 1.2 &  & K & 55 \\
 & Ba-Bb & 1993 & 330 & 0.4 & 0.267 & 0.4 & 2.0647 & spkdet & 55 & 0.35 &  & K & 55 \\
00322970+6714080 & AaAb-B & 1923 & 99 &  & 2.26 &  & 22.511 & phodet & 86 & 2.5 &  & V & 21 \\
 & Aa-Ab & 1989 & 43 & 3 & 0.451 & 0.02 & 4.4922 & spkdet & 66 & 1.94 & 0.06 & K & 66 \\
LTT 10301 & A-B & 1960 & 315 &  & 1 &  & 14.986 & phodet & 58 & 0.3 &  & B & 58 \\
00582789-2751251 & A-B &  &  &  &  &  & 0.2381 & SB1 & 23 &  &  &  &  \\
01023895+6220422 & A-B & 1999 & 76 &  & 293.05 &  & 2889.4 & visdet & 75 & 2.35 & 0.03 & K & 75 \\
01053732+2829339 & A-B &  &  &  &  &  & 0.133 & SB2 & 3 &  &  &  &  \\
\enddata
\tablecomments{Table \ref{table:known_companions} is published in its entirety in the machine-readable format. A portion is shown here for guidance regarding its form and content.}
\tablecomments{The codes for the techniques and instruments used to detect and resolve systems are: AO det—detection via adaptive optics; astdet—detection via astrometric perturbation, companion often not detected directly; astorb—orbit from astrometric measurements; CCDdet—detection via CCD or other two-dimensional electronic imaging; lkydet—detection via lucky imaging; micdet-detection via a micrometry technique; occdet-detection via occulation; phodet—detection via a photographic technique; radvel—detection via radial velocity, but no SB type indicated; SB (1, 2, 3)—spectroscopic multiple, either single-lined, double-lined, or triple-lined; spcdet-detection via space-based technique; spkdet—detection via speckle interferometry; visdet—detection via wide-field CCD or other two-dimensional electronic imaging.}
\tablerefs{(1) \citet{Aitken1899AN....150..113A}; (2) \citet{Balega1999AAS..140..287B}; (3) \citet{Baroch2018AA...619A..32B}; (4) \citet{Benedict2000AJ....120.1106B}; (5) \citet{Beuzit2004AA...425..997B}; (6) \citet{Blazit1987AAS...71...57B}; (7) \citet{Bowler2015ApJS..216....7B}; (8) \citet{Burnham1891AN....127..289B}; (9) \citet{Burnham1891AN....127..369B}; (10) \citet{Burnham1894PLicO...2....1B}; (11) \citet{Burnham1913mpms.book.....B}; (12) \citet{Clark2022AJ....164...33C}; (13) \citet{Clark2023RNAAS...7..206C}; (14) \citet{Cortes-Contreras2017AA...597A..47C}; (15) \citet{Dahn1976PUSNO..24c...1D}; (16) \citet{Delfosse1999AA...344..897D}; (17) \citet{Delfosse1999AA...350L..39D}; (18) \citet{DuquennoyMayor1988AA...200..135D}; (19) \citet{ESA1997ESASP1200.....E}; (20) \citet{Espin1920MNRAS..80..329E}; (21) \citet{EspinMilburn1926MNRAS..86..131E}; (22) \citet{Forrest1988ApJ...330L.119F}; (23)
\citet{Fouque2018MNRAS.475.1960F}; (24) \citet{Franz1998AJ....116.1432F}; (25) \citet{Gaia2023AA...674A...1G}; (26) \citet{Giclas1971lpms.book.....G}; (27) \citet{Gili2022AN....34324008G}; (28) \citet{GizisReid1996AJ....111..365G}; (29)   \citet{Hartkopf1994AJ....108.2299H}; (30) \citet{Heintz1987ApJS...65..161H}; (31) \citet{Heintz1993AJ....105.1188H}; (32) \citet{Henry1997AJ....114..388H}; (33) \citet{Henry1999ApJ...512..864H}; (34) \citet{HerbigMoorhead1965ApJ...141..649H}; (35) \citet{Hertzsprung1909AN....180...39H}; (36) \citet{Horch2004AJ....127.1727H}; (37) \citet{Horch2010AJ....139..205H}; (38) \citet{Horch2012AJ....143...10H}; (39) \citet{Hough1899AN....149...65H}; (40) \citet{Hussey1904LicOB...2..180H}; (41) 
\citet{Janson2012ApJ...754...44J}; (42) \citet{Janson2014ApJ...789..102J}; (43) \citet{Jeffers2018AA...614A..76J}; (44) \citet{Jodar2013MNRAS.429..859J}; (45) 
\citet{Joy1942PASP...54...33J};(46) \citet{Joy1947ApJ...105...96J}; (47) \citet{Kuiper1934PASP...46..360K}; (48) \citet{Kuiper1936ApJ....84..478K}; (49) \citet{Kuiper1936ApJ....84R.359K}; (50) \citet{Kuiper1942ApJ....96..315K}; (51) \citet{Kuiper1943ApJ....97..275K}; (52) \citet{Lamman2020AJ....159..139L}; (53) \citet{Lau1911AN....189..197L}; (54) \citet{Leinert1986AA...164L..29L}; (55) \citet{Leinert1994AA...291L..47L}; (56) \citet{Luyten1969PMMin..21....1L}; (57) \citet{Luyten1972PMMin..29....1L}; (58) \citet{Luyten1977PMMin..50....1L}; (59) \citet{Luyten1979nltt.book.....L}; (60) \citet{Luyten1997yCat.1130....0L}; (61) \citet{Luyten1998yCat.1087....0L}; (62) \citet{Malo2014ApJ...788...81M}; (63) \citet{Martin2000ApJ...529L..37M}; (64) \citet{McCarthyHenry1987ApJ...319L..93M}; (65) \citet{McCarthy1988ApJ...333..943M}; (66) \citet{McCarthy1991AJ....101..214M}; (67) \citet{Montagnier2006AA...460L..19M}; (68) \citet{OsvaldsOsvalds1959AJ.....64..265O};(69) \citet{ReidGizis1997AJ....113.2246R}; (70)
\citet{ReinersBasri2010ApJ...710..924R}; (71) \citet{Reuyl1941PASP...53..119R}; (72) \citet{Richichi1996AA...309..163R}; (73) 
\citet{Riedel2010AJ....140..897R}; (74) \citet{Rossiter1955POMic..11....1R}; (75) \citet{Skrutskie2006AJ....131.1163S}; (76) 
\citet{Sperauskas2019AA...626A..31S}; (77) \citet{Struve1837AN.....14..249S}; (78) \citet{Tokovinin2010AJ....139..743T}; (79) \citet{TomkinPettersen1986AJ.....92.1424T}; (80) \citet{vanBiesbroeck1961AJ.....66..528V}; (81) \citet{vanBiesbroeck1974ApJS...28..413V}; (82) \citet{vandenBos1937CiUO...98..362V}; (83) \citet{vandenBos1950CiUO..109Q.371V}; (84) \citet{vandenBos1951CiUO..111...13V}; (85) \citet{Vrijmoet2022AJ....163..178V}; (86) \citet{Vyssotsky1927PA.....35..213V}; (87) \citet{Ward-Duong2015MNRAS.449.2618W}; (88) \citet{Wendell1913AnHar..69...99W}; (89) \citet{Wilson1954AJ.....59..132W}; (90) \citet{Winters2017AJ....153...14W}; (91) 
\citet{Winters2020AJ....159..290W}; (92) \citet{Wirtanen1941PASP...53..340W}; (93) \citet{Worley1962AJ.....67..403W}.}
\end{deluxetable*}

In total, we find that \multiplefifteen{} of the \primaryfifteen{} POKEMON primaries within 15 pc host M-dwarf companions known to the literature. Of these, \doublefifteen{} are double ($N_D$), \triplefifteen{} are triple ($N_T$), \quadruplefifteen{} are quadruple ($N_{Qd}$), and \quintuplefifteen{} is quintuple ($N_{Qn}$). \singlefifteen{} of the POKEMON primaries within 15 pc are therefore not known to host stellar companions ($N_S$). We can calculate the multiplicity and companion rates using these values and the equations defined in \citet{Winters2019AJ....157..216W}:

\begin{equation}
    MR = 100 * \frac{N_D + N_T + N_{Qd} + N_{Qn}}{N_S + N_D + N_T + N_{Qd} + N_{Qn}}
\end{equation}

\begin{equation}
    CR = 100 * \frac{N_D + 2N_T + 3N_{Qd} + 4N_{Qn}}{N_S + N_D + N_T + N_{Qd} + N_{Qn}}
\end{equation}

The multiplicity rate is simply defined as the percentage of systems that are multiple. In comparison, the companion rate is defined as the average number of companions per primary in the sample, so it takes into account whether the systems are binary, trinary, or of even higher orders.

We calculate the uncertainty on our rates using the analytic formula for the sample standard deviation of a binomial distribution

\begin{equation}
    \sigma_n = \sqrt{\frac{p (1 - p)}{n}}
\end{equation}

\noindent where $n$ is the number of experiments (the number of primaries) and $p$ is the probability of success (the number of multiples divided by the number of primaries).

Using these equations, we find a stellar multiplicity rate of \multiplicityrate{}, and a stellar companion rate of \companionrate{}. We also find that the ratios of singles:doubles:triples:higher-order systems is \singlefifteen{}:\doublefifteen{}:\triplefifteen{}:\quintuplefifteen{}, corresponding to 76.5:18.7:4.6:0.2\%. For comparison, \citet{Winters2019AJ....157..216W} found ratios of 76:20:3:0.3\% for their sample of M dwarfs within 25 pc. 104 of the targets in the 15-pc POKEMON sample are not in the \citet{Winters2019AJ....157..216W} sample, and yet the ratios are quite similar. In contrast, \citet{Raghavan2010ApJS..190....1R} found ratios of 56:33:8:3\% for singles:doubles:triples:higher-order systems for companions (including brown dwarfs) in a sample of 454 solar-type stars. This result re-emphasizes that more massive stars are more likely to be higher-order systems than our less-massive neighbors.

\begin{figure*}
    \centering
    \includegraphics[width=\textwidth]{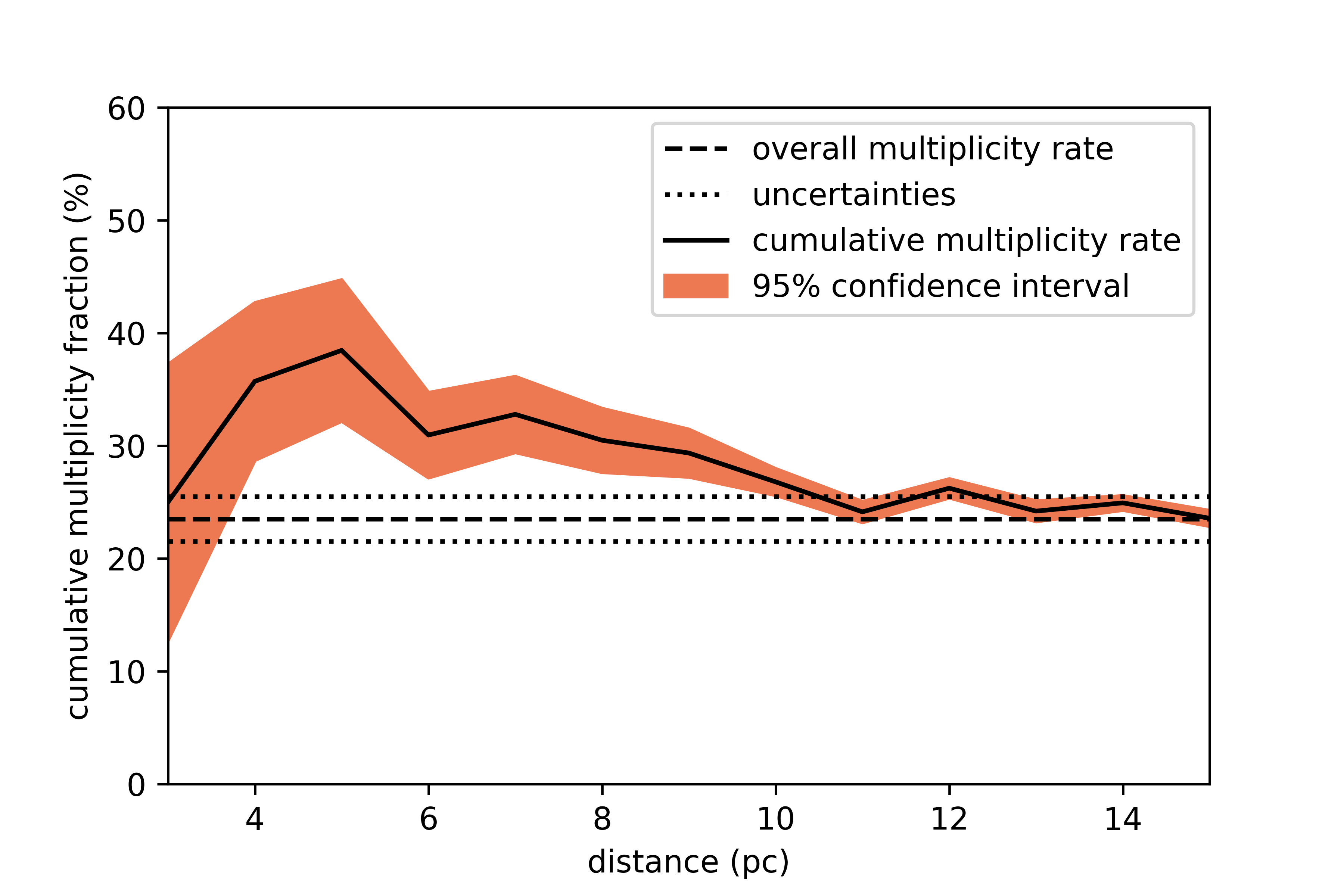}
    \caption{Cumulative stellar multiplicity rate as a function of distance, binned by one parsec. The 95\% Poisson confidence interval is shown. We also show the overall stellar multiplicity rate (and uncertainties) for comparison. No corrections have been applied for undetected companions. Although the cumulative multiplicity rate is higher at smaller distances due to a smaller number of stars in those bins, we do find that the cumulative multiplicity rate stays roughly flat from 11 pc outwards; the drop-off from 11 to 15 pc is 0.6\%, which we consider negligible. Since the cumulative multiplicity rate remains roughly flat in these bins, this indicates that we are not missing multiples at these distances.}
    \label{fig:multiplicity_by_distance}
\end{figure*}

In Figure \ref{fig:multiplicity_by_distance}, we plot the cumulative stellar multiplicity rate as a function of distance. Although the cumulative multiplicity rate is higher at smaller distances due to a smaller number of stars in those one-parsec bins, we do find that the cumulative multiplicity fraction stays within the uncertainties of the overall multiplicity rate from 11 pc outwards; the dropoff from 11 to 15 pc is 0.6\%, which we consider negligible. Since the cumulative multiplicity function remains roughly flat in these bins, this indicates that we are likely not missing multiples.

As noted previously, these multiplicity rates are restricted to include only stellar companions, and not brown dwarfs. It is therefore important to explore whether this exclusion of substellar companions may impact our calculated multiplicity rate. We note that lower-mass binaries tend to have mass ratios closer to unity, and surveys to date have found few brown dwarfs around small stars using radial velocity measurements \citep{Tokovinin1992A&A...256..121T, Pass2023AJ....166...11P}, high-resolution imaging \citep{Dieterich2012AJ....144...64D}, or astrometry \citep[e.g.,][]{Henry2018AJ....155..265H}. Furthermore, \citet{Winters2019AJ....157..216W} and \citet{Raghavan2010ApJS..190....1R} find quite low brown dwarf companion rates of $1.3\pm0.3\%$ and $1.5\pm0.6\%$, respectively. We therefore conclude that the exclusion of brown dwarf companions has little impact on our multiplicity rate calculations.

\section{The Separation Gap} \label{sec:stellar_multiple_gap}

As expected, close-in stellar companions affect multiple facets of planet formation, as well as the dynamics and architectures of planetary systems. \citet{Cieza2009ApJ...696L..84C} analyzed young stars with protoplanetary disks, and showed that disk-hosting multiples have a very different projected separation distribution as compared to non-disk-hosting multiples. Systems with close-in stellar multiples ($a<40$ au) are half as likely to retain a disk than more widely-separated multiples.

More recently, studies of data from Kepler, K2, and TESS have shown that when we detect stellar companions to planet-hosting FGK stars, the companions have projected separation distributions that peak at larger values than what have been found for field stars \citep[e.g.,][]{Kraus2012ApJ...745...19K, Bergfors2013MNRAS.428..182B, Wang2014ApJ...791..111W, Kraus2016AJ....152....8K, Ziegler2020AJ....159...19Z, Howell2021AJ....161..164H, MoeKratter2021MNRAS.507.3593M}.

Complementing these works, a survey combining radial velocity and imaging observations of nearby FGK stars found that the frequency of giant planets around single stars and in wide ($\gtrsim 100$ au) binaries was nearly identical ($\approx 20\%$), while the frequency of giant planets in close ($\lesssim 100$ au) binaries was nearly 0\% \citep{Hirsch2021AJ....161..134H}. This is likely the result of the close-in stellar companions affecting and perhaps inhibiting the planet formation process \citep[e.g.,][]{Dupuy2016ApJ...817...80D}. While these works focused on solar-type (FGK) stars, \citet{Clark2022AJ....163..232C} saw this in a sample of M-dwarf TESS Objects of Interest (TOIs) as well. In that work, the frequency of stellar companions around TOI-hosting M dwarfs was found to be similar to the companion rate for field stars, but the stellar companions were found at much larger separations.

The volume-limited, 15-pc POKEMON sample enables an exploration of this idea for the M dwarfs in a systematic manner. The projected separation distribution for the multiples in the 15-pc POKEMON sample is roughly Gaussian with a peak at 6.57 au ($\sigma_{\log a}=1.3$, $SE_{\log a}=0.11$). Specifically within 10 pc, we find a peak at 4.95 au ($\sigma_{\log a}=1.4$, $SE_{\log a}=0.18$). This is in good agreement with the recent survey of M dwarfs for wide ($\gtrsim 2\arcsec$) companions \citep{Winters2019AJ....157..216W}, which found a peak at 20 au for their sample within 25 pc, and at 4 au for their sample within 10 pc. However, when we explore the projected separation distribution for the POKEMON multiples that host planets, we find that it is quite different from the \citet{Winters2019AJ....157..216W} distribution.

There are seven POKEMON primaries within 15 pc that are known to host both stellar and planetary companions (Table \ref{table:known_planets}). GJ 15 A and LTT 1445 A host two planets each, so in total there are eight known planets. These systems were identified via a search of the Exoplanet Archive\footnote{https://exoplanetarchive.ipac.caltech.edu} for known planets in multi-star systems and a cross-match with our sample. Interestingly, only one planetary system was detected via the transit method, which has been used to identify the vast majority of known planets. 419 of the \primaryfifteen{} targets in the 15-pc POKEMON sample have been observed by TESS, and of these, 397 have a TESS magnitude brighter than 13, meaning they have been searched by the TESS project's quick-look pipeline for planets. This means that in total, 87\% of the 15-pc POKEMON sample has been searched for transiting planets. There are eight TOIs in the 15-pc POKEMON sample: 256, 455, 562, 1796, 1827, 4481, 4599, and 6251. However, only one of these TOIs (455, or LTT 1445 A) was discovered in a system with stellar companions.

\startlongtable
\begin{deluxetable*}{cccc}
\tablecaption{POKEMON primaries within 15 pc that host both planetary and stellar companions
\label{table:known_planets}}
\tablehead{\colhead{Planet} & \colhead{Discovery Technique} & \colhead{Reference} & \colhead{Stellar Companion Separation} \\
\colhead{} & \colhead{} & \colhead{} & \colhead{(au)}}
\startdata
 GJ 15 A b & Radial Velocity & \citet{Howard2014ApJ...794...51H} & 138.33 \\
 GJ 15 A c & Radial Velocity & \citet{Pinamonti2018AA...617A.104P} & 138.33 \\
 GJ 49 b & Radial Velocity & \citet{Perger2019AA...624A.123P} & 2889.4 \\
 LTT 1445 A b & Transit & \citet{Winters2019AJ....158..152W} & 22.563 \\
 LTT 1445 A c & Transit & \citet{Winters2022AJ....163..168W} & 22.563 \\
 Ross 458 c & Imaging & \citet{Burgasser2010ApJ...725.1405B} & 5.466 \\
 HD 147379 b & Radial Velocity & \citet{Reiners2018AA...609L...5R} & 691.99 \\
 HD 180617 b & Radial Velocity & \citet{Kaminski2018AA...618A.115K} & 437.71 \\
 GJ 896 A b & Astrometry & \citet{Curiel2022AJ....164...93C} & 21.921 \\
 \enddata
\end{deluxetable*}

We note that the presence of a blended stellar companion can prevent the detection of transiting small planets by diluting the transit depth below the detection threshold of TESS \citep{Lester2021AJ....162...75L}. The approximate transit depth limit for TESS is $\sim100$ ppm; after correction for an equal-brightness companion, the minimum transit depth would be $\sim200$ ppm, translating to a planet that is $1.4\times$ larger than originally thought. At 200 ppm, the minimal detectable planet would be $0.9 R_{\oplus}$ around an M0V star, and $0.2 R_{\oplus}$ around an M9V star (E. Mamajek, private communication). Thus, there could be very small planets hidden in the sample beyond the detection limits of TESS.

Additionally, we note that short-period transiting planets discovered by TESS do not represent the entire range of planets that may exist in the 15-pc POKEMON sample. While 87\% of the 15-pc sample was observed by TESS, transiting planets represent only about 10\% of the expected total number of planets. However, radial velocity searches for planets orbiting these stars are substantially less complete and/or systematic. Many of these systems have not been searched via the radial velocity method due to the faintness of the host star and/or the presence of a stellar companion. A full completeness and reliability analysis of the radial velocity surveys is beyond the scope of this work, but we note that non-transiting planets could be hidden in the sample. Nonetheless, the lack of planets we find in the 15-pc POKEMON sample is consistent with the radial velocity+imaging survey of nearby (within 300 pc) FGK stars, where Neptune-sized and larger planets in short orbital periods were rarely found in systems with close-in stellar companions \citep{Hirsch2021AJ....161..134H}.

\begin{figure*}
    \centering
    \includegraphics[width=\textwidth]{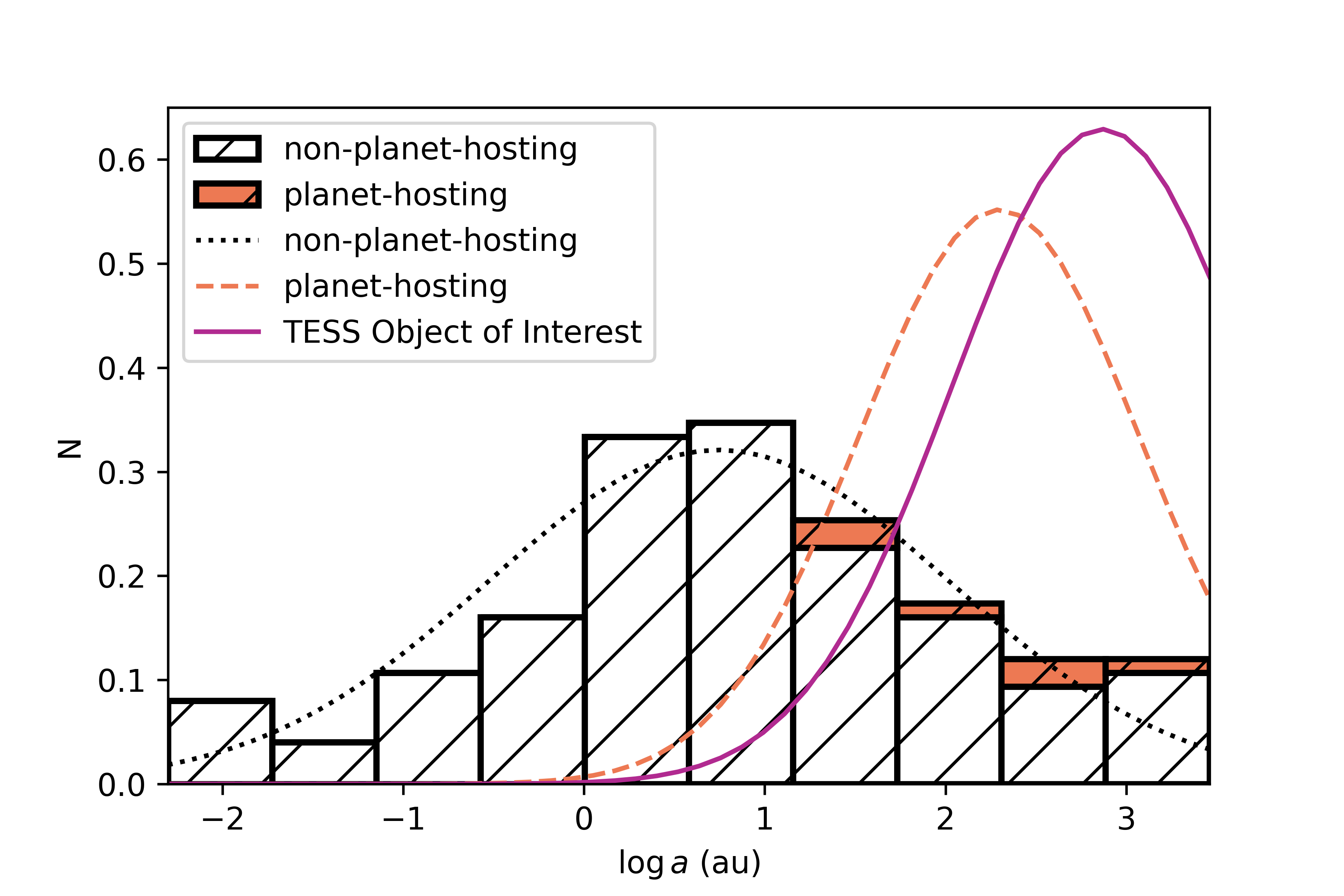}
    \caption{Projected separation distributions for various populations of M-dwarf multiples. The histogram shows the distribution of projected separations for the targets in the 15-pc POKEMON sample that are known to be multiple, separated by planet-host (orange) and non-planet-host (white) status. The dotted, black line is a Gaussian fit to the projected separations of the non-planet-hosting POKEMON multiples, with a peak at \noplanetpeak{} au ($\sigma_{\log a}=1.3$, $SE_{\log a}=0.11$). The dashed, orange line is a Gaussian fit to the projected separations of the planet-hosting POKEMON multiples, with a peak at \planetpeak{} au ($\sigma_{\log a}=0.77$, $SE_{\log a}=0.32$). The solid, purple line is a Gaussian fit to the projected separations of the TOIs from \citet{Clark2022AJ....163..232C}, with a peak at \toipeak{} au ($\sigma_{\log a}=0.83$, $SE_{\log a}=0.25$). We note that we use Gaussian fits to compare our distribution of projected separations with other studies of multiplicity \citep[e.g.,][]{Raghavan2010ApJS..190....1R, Winters2019AJ....157..216W}; we do not use this to indicate that the distribution of stellar multiples is Gaussian, but rather to show that the peaks of the distributions are shifted. The two planet-hosting distributions have peaks at relatively consistent values, whereas the peaks of the planet-hosting and TOI distributions are larger than that of the non-planet-hosting distribution by more than an order of magnitude. This result indicates that the presence of a stellar companion impacts the formation and long-term stability of any planets in the system.}
    \label{fig:proj_sep_gap}
\end{figure*}

With these caveats noted, in Figure \ref{fig:proj_sep_gap} we plot the histogram of projected separations for the multiples in the 15-pc POKEMON sample, as well as Gaussian fits to the planet-hosting and non-planet-hosting distributions, and TOI distributions from \citet{Clark2022AJ....163..232C}. When the non-planet-hosting and planet-hosting samples are assessed separately, the non-planet-hosting POKEMON multiples have a projected separation distribution that peaks at \noplanetpeak{} ($\sigma_{\log a}=1.3$, $SE_{\log a}=0.11$), while the planet-hosting POKEMON multiples have a peak at \planetpeak{} au ($\sigma_{\log a}=0.77$, $SE_{\log a}=0.32$). These peaks and standard deviations are determined when the log10-separations are fit with a normal Gaussian distribution, and are taken to be the means of the distributions. We note that we consider the Ross 458 system non-planet-hosting, as its `planet’ is a circumbinary T dwarf located $\sim100\arcsec$ from the M-dwarf pair.

While only a small number of POKEMON multiples host known planets, the planet-hosting distribution is in reasonable agreement with the results from the recent multiplicity survey of M-dwarf TOIs \citep{Clark2022AJ....163..232C}. After removing false positives, the peak of the projected separation distribution for the M-dwarf TOIs is found to be \toipeak{} au ($\sigma_{\log a}=0.83$, $SE_{\log a}=0.25$).

We also break down the projected separations of the stellar multiples by planet-host status in Table \ref{table:planet_distributions}, where we note that the breakdown of close-in and wide companions is $\sim5:1$ for the non-planet-hosting multiples, but reversed to $\sim1:5$ for the planet-hosting multiples.

\startlongtable
\begin{deluxetable*}{ccc}
\tablecaption{Projected separations of stellar multiples by planet-hosting status
\label{table:planet_distributions}}
\tablehead{\colhead{} & \colhead{$a<100$ au}  & \colhead{$a>100$ au}}
\startdata
Non-Planet-Hosting POKEMON Targets & 102 & 22 \\
Planet-Hosting POKEMON Targets  & 2 & 4 \\
M-Dwarf TESS Objects of Interest & 2 & 9 \\
 \enddata
\end{deluxetable*}

The number of planet-hosting systems with stellar companions in both the POKEMON (6) and TOI (11) samples is relatively small, but it is telling the two planet-hosting distributions have peaks at relatively consistent values, whereas the peaks of the planet-hosting and non-planet-hosting distributions differ by more than an order of magnitude. Every planet-hosting binary -- besides the circumbinary planet -- has a companion separation that is larger than the peak of the non-planet-hosting separation distribution. Additionally, an Anderson-Darling test on the planet-hosting and non-planet-hosting POKEMON multiple distributions yields a p-value of 0.003, indicating that the two samples are likely drawn from different parent distributions. Furthermore, a Kolmogorov–Smirnov (K-S) test on the planet-hosting and non-planet-hosting POKEMON multiple distributions yields a p-value of 0.003. When the non-planet-hosting POKEMON multiples are compared to M-dwarf TESS Objects of Interest from \citet{Clark2022AJ....163..232C}, the K-S test has the same result ($\text{p-value}=7\sci{e-7}$). However, a K-S test on the planet-hosting POKEMON multiples and the M-dwarf TOIs shows that the two distributions likely stem from the same parent distribution ($\text{p-value}=0.5$).

In order to determine the statistical significance of the difference between the peaks of the planet-hosting and non-planet-hosting samples, we added the standard errors of the planet-hosting and non-planet-hosting samples in quadrature to estimate the uncertainty in the difference; we find a value of 2.17 au. This means that the peaks of the planet-hosting and non-planet-hosting differ by over $88\sigma$, a highly-significant result.

We also carried out a Monte Carlo analysis to investigate this result further. We drew six random samples -- representing the six planet-hosting multiples -- from the projected separation distribution for the POKEMON multiples one thousand times. We then fit a Gaussian distribution to each draw to calculate the mean. This mean was equal to or larger than the mean of the planet-hosting POKEMON multiple distribution in only one out of one thousand cases. This result indicates that it is very unlikely that the highly-skewed peak of the planet-hosting POKEMON multiples is the result of random chance.

These results represent statistically-significant evidence for a correlation between binary separation and the presence of a planetary companion in M-dwarf systems. Of course, observational biases may exist in our samples; for instance, the low-separation side of the histogram could be affected by still-missing companions. And as noted, the POKEMON sample has not been uniformly-searched for planets, and there are likely (as-of-now) unknown planets orbiting stars in the non-planet-hosting portion of the sample. Additionally, the planetary content of the secondary stars in these systems is relatively unknown. It would therefore be of interest to search both components of systems with wider companions ($\gtrsim 100$ au) for planets to investigate whether the tail towards larger projected separations that we find in the POKEMON multiple distribution is produced by planet-hosting systems, and to investigate whether the planets that orbit secondary stars differ from those that orbit primaries.

\section{Conclusions and Future Work} \label{sec:conclusions}

We have carried out the POKEMON speckle survey of nearby M dwarfs using the DSSI and NESSI instruments on the 4.3-meter LDT and the 3.5-meter WIYN telescope, respectively. Using these instruments, we have imaged \pokemonnum{} targets, \primaryfifteen{} of which are in our volume-limited, 15-pc sample of M-dwarf primaries. We find that the stellar multiplicity rate of M dwarfs within 15 pc is \multiplicityrate{}, and that the companion rate is \companionrate{}.

We also investigated how the presence of stellar companions may impact the properties of any planetary companions in the system. We find that peaks of the projected separation distributions for non-planet-hosting and planet-hosting multiples differ by over an order of magnitude (\noplanetpeak{} and \planetpeak{} au, respectively). This result is consistent with the recent survey of M-dwarf TOIs, which have a peak at \toipeak{} au \citep{Clark2022AJ....163..232C}. We therefore conclude that planet-hosting multiples have larger projected separations than non-planet-hosting multiples, indicating that the presence of a stellar companion impacts the formation and long-term stability of any planets in the system. Planets orbiting M dwarfs are favorable targets for atmospheric characterization via transmission spectroscopy, so understanding the occurrence rate and properties of stellar companions to M dwarfs is critical to current missions such as the James Webb Space Telescope \citep{Gardner2006SSRv..123..485G}, as well as upcoming missions such as Ariel \citep{Tinetti2018ExA....46..135T} and the Habitable Worlds Observatory.

We are continuing to follow up the POKEMON targets with the Quad-camera Wavefront-sensing Six-channel Speckle Interferometer \citep{Clark2020SPIE11446E..2AC}. We are also currently obtaining homogenous spectral types for all targets in the sample, and in a forthcoming publication we will present the stellar multiplicity rate of M dwarfs within 15 pc calculated by spectral subtype through M9V for the first time.

\clearpage

\begin{acknowledgments}

We are so grateful to the multitude of folks who have contributed to the POKEMON survey, and in particular to Brian Mason and Frederick Hahne. We also thank the anonymous reviewer, Davy Kirkpatrick, and Eric Mamajek for their contributions to this manuscript.

This research was carried out at the Jet Propulsion Laboratory, California Institute of Technology, under a contract with the National Aeronautics and Space Administration (80NM0018D0004). This research was supported by NSF Grant No.~AST-1616084 awarded to GTvB and NASA Grant 18-2XRP18\_2-0007 awarded to DRC.

This work presents results from the European Space Agency (ESA) space mission Gaia
. Gaia data are being processed by the Gaia Data Processing and Analysis Consortium (DPAC). Funding for the DPAC is provided by national institutions, in particular the institutions participating in the Gaia MultiLateral Agreement (MLA). The Gaia mission website is \url{https://www.cosmos.esa.int/gaia}. The Gaia archive website is \url{https://archives.esac.esa.int/gaia}.

This work has used data products from the Two Micron All Sky Survey \citep{https://doi.org/10.26131/irsa2}, which is a joint project of the University of Massachusetts and the Infrared Processing and Analysis Center at the California Institute of Technology, funded by NASA and NSF.

This research has made use of the Exoplanet Follow-up Observation Program (ExoFOP, DOI:10.26134/ExoFOP5) website, which is operated by the California Institute of Technology, under contract with the National Aeronautics and Space Administration under the Exoplanet Exploration Program.

This research has made use of the NASA Exoplanet Archive, which is operated by the California Institute of Technology, under contract with the National Aeronautics and Space Administration under the Exoplanet Exploration Program.

Information was collected from several additional large database efforts: the Simbad database and the VizieR catalogue access tool, operated at CDS, Strasbourg, France; NASA's Astrophysics Data System; and the Washington Double Star Catalog maintained at the US Naval Observatory.

© 2023. California Institute of Technology. Government sponsorship acknowledged.

\end{acknowledgments}

%

\vspace{5mm}
\facilities{Exoplanet Archive}


\software{Astropy \citep{Astropy2013}, Astroquery \citep{Astroquery2019AJ....157...98G}, IPython \citep{IPython2007}, Matplotlib \citep{Matplotlib2007}, NumPy \citep{NumPy2020}, Pandas \citep{Pandas2010}, SciPy \citep{SciPy2020}}

\clearpage





\bibliography{references}{}
\bibliographystyle{aasjournal}



\end{document}